\documentclass[pdflatex,sn-mathphys-num]{sn-jnl}

\usepackage{graphicx}%
\usepackage{multirow}%
\usepackage{amsmath,amssymb,amsfonts}%
\usepackage{amsthm}%
\usepackage{mathrsfs}%
\usepackage[title]{appendix}%
\usepackage{xcolor}%
\usepackage{textcomp}%
\usepackage{manyfoot}%
\usepackage{booktabs}%
\usepackage{algorithm}%
\usepackage{algorithmicx}%
\usepackage{algpseudocode}%
\usepackage{listings}%

\usepackage{quantikz}
\usepackage{xurl}
\usepackage{hyperref}

\theoremstyle{thmstyleone}%
\theoremstyle{thmstyletwo}%

\theoremstyle{thmstylethree}%

\begin{document}

\title[Article Title]{Beauty and charm jet identification with end-to-end training on a superconducting quantum processor}


\author[1]{\fnm{Lucio} \sur{Anderlini}}
\author[2,3]{\fnm{Pietro} \sur{Asinari}}
\author[3]{\fnm{Gianluca} \sur{Bertaina}}
\author[3]{\fnm{Davide} \sur{Calonico}}
\author[3]{\fnm{Marco} \sur{Co\"isson}}
\author[4]{\fnm{Alessio} \sur{Gianelle}}
\author[4,5]{\fnm{Donatella} \sur{Lucchesi}}
\author[1]{\fnm{Giovanni} \sur{Passaleva}}
\author[2]{\fnm{Matteo Maria} \sur{Piredda}}

\author*[1]{\fnm{Lorenzo} \sur{Sestini}}
\email{lorenzo.sestini@fi.infn.it}

\author[1,6]{\fnm{Michele} \sur{Veltri}}
\author[4,5]{\fnm{Davide} \sur{Zuliani}}

\affil[1]{
  \orgname{Istituto Nazionale di Fisica Nucleare (INFN), Sezione di Firenze},
  \orgaddress{\city{Florence}, \country{Italy}}
}

\affil[2]{
  \orgdiv{Dipartimento Energia},
  \orgname{Politecnico di Torino},
  \orgaddress{\city{Turin}, \country{Italy}}
}

\affil[3]{
  \orgname{Istituto Nazionale di Ricerca Metrologica (INRiM)},
  \orgaddress{\city{Turin}, \country{Italy}}
}

\affil[4]{
  \orgname{Istituto Nazionale di Fisica Nucleare (INFN), Sezione di Padova},
  \orgaddress{\city{Padua}, \country{Italy}}
}

\affil[5]{
  \orgname{Universit\`a di Padova},
  \orgaddress{\city{Padua}, \country{Italy}}
}

\affil[6]{
  \orgname{Universit\`a di Urbino},
  \orgaddress{\city{Urbino}, \country{Italy}}
}


\abstract{The identification of heavy-flavor jets, i.e.\ collimated streams of particles originating from the fragmentation of beauty and charm quarks produced in proton–proton collisions, represents one of the most challenging classification tasks in measurements at Large Hadron Collider experiments. In recent years, advanced machine learning techniques have been successfully employed to address this problem, while quantum machine learning has emerged as a promising alternative approach. However, the high dimensionality of the problem, typically involving hundreds of observables, raises an open question: whether the current generation of quantum computers, characterized by limited qubit count, can effectively tackle such complex tasks.
In this work, a variational quantum circuit for the classification of $b$- and $c$-jets at the LHCb experiment is implemented on a 5-qubit superconducting quantum computer manufactured by IQM and installed at the Politecnico di Torino. To the best of our knowledge, this is the first demonstration of a quantum algorithm for jet identification using simulated data from an LHC experiment in which all stages of the machine-learning workflow, including training and testing, are performed on real quantum hardware. The results demonstrate that the impact of hardware-induced quantum noise does not significantly degrade the performance. A framework for the evaluation of the training time on the quantum hardware and for the circuit optimization is also proposed.
These findings pave the way for future applications on larger quantum devices based on superconducting qubits, featuring a larger number of qubits enabling the encoding of information beyond the reach of classical simulation and providing a step toward investigating whether quantum advantage may eventually be achievable in high-energy physics classification tasks.}




\maketitle

\section{Introduction}\label{sec1}

Artificial intelligence (AI) is widely used to solve a variety of tasks in accelerator-based particle physics \cite{hep_ai}. One of the most prominent applications is the identification of hadronic jets produced in proton–proton collisions at the Large Hadron Collider (LHC) \cite{jet_ai}. Jets are collimated streams of particles originating from the fragmentation and hadronization of quarks and gluons produced in high-energy collisions. In particular, jets originating from beauty ($b$) and charm ($c$) quarks are of special interest, as they appear in the final states of many processes used to test the Standard Model (SM) of particle physics.
Both $b$- and $c$-hadrons emerging from jet fragmentation have comparable lifetimes ($\tau_b \sim 1.5$ ps and $\tau_c \sim 0.2{-}1$ ps). Moreover, $b \to c$ transitions are dominant, leading to similar topological features between $b$- and $c$-jets. For these reasons, the discrimination between $b$- and $c$-jets, commonly referred to as heavy-flavor tagging, represents a challenging classification task. To address it, LHC experiments, namely ATLAS~\cite{ATLAS:2008}, CMS~\cite{CMS:2008}, LHCb~\cite{lhcb}, and ALICE~\cite{ALICE:2008}, have developed sophisticated tagging strategies based on hundreds of reconstructed observables, which are then processed using advanced AI-based algorithms.

In this context, quantum machine learning (QML) has emerged as a promising paradigm for jet identification tasks \cite{Chen:2025, qml_tagging, gluon_quark_tagging}. Several feasibility studies based on variational quantum circuits (VQCs) have been proposed; however, their performance is generally comparable to that of classical methods such as neural networks \cite{gnn} and boosted decision trees (BDTs) \cite{bdt}. A key limitation of current approaches lies in the restricted number of qubits available in existing quantum processors and simulators. 
In practice, the computational resources required for exact statevector simulation grow exponentially with the number of qubits, eventually making the classical simulation of sufficiently large quantum circuits computationally demanding.
Depending on the chosen encoding strategy, the limited number of available qubits may constrain the dimensionality of the input feature space.
Consequently, realistic jet tagging problems involving hundreds of observables remain out of reach for current quantum simulators. It is expected that only when quantum processors with on the order of 100 qubits or more become routinely available will it be possible to encode more realistic representations of the full jet feature space and perform more comprehensive comparisons between quantum and classical approaches. Such studies will be essential for investigating the possibility of quantum advantage~\cite{Biamonte:2017}, understood here as a demonstrable computational or performance benefit of a quantum approach over relevant classical methods for the same classification task. One possible source of such an advantage may be found in the ability of quantum circuits to exploit non-classical correlations, which could be relevant in processes governed by quantum field theory such as fragmentation and hadronization \cite{hadronization}. Another advantage is a possible better scaling of energy consumption with increasing model accuracy \cite{qml_energy}. Recently, it has been claimed that an on-the-fly QML algorithm could vastly outperform analogously accurate classical machine learning methods~\cite{zhao2026exponentialquantumadvantageprocessing}. For specific classification and dimensional-reduction tasks, this approach was shown to require only polylogarithmic quantum memory, while classical algorithms achieving the same prediction performance require exponentially larger memory resources.


Moreover, open questions remain regarding when such large-scale quantum hardware will become reliably available for scientific applications, and whether the associated noise levels will allow for meaningful training and inference. In particular, VQCs require repeated evaluations on quantum hardware during training, making them highly sensitive to noise and statistical fluctuations. Training VQCs on real quantum processors for realistic, large-scale classification tasks remains highly resource-demanding with currently available quantum hardware~\cite{Chen:2025, gluon_quark_tagging}, even at small scales, due to limited availability of quantum processing unit (QPU) time and associated costs.

In the Italian quantum computing landscape, several commercial and experimental quantum hardware platforms are emerging and becoming increasingly accessible to research institutions, marking a significant step toward the development of a national quantum computing ecosystem. A notable example is the 54-qubit superconducting quantum computer developed by IQM~\cite{IQM:2024}. The system has been acquired by CINECA, one of Italy's leading high-performance computing centers providing advanced computing resources to universities and research institutions~\cite{CINECA:IQM54}, and is currently being deployed. This system is integrated within a high-performance computing environment, enabling the exploration of hybrid quantum–classical workflows and providing access to Noisy Intermediate-Scale Quantum (NISQ) hardware~\cite{Preskill:2018}. In addition, a smaller-scale 5-qubit system from IQM has been installed at the Politecnico di Torino~\cite{PoliTO:web}, with the contribution of Fondazione LINKS and Istituto Nazionale di Ricerca Metrologica (INRiM)~\cite{INRiM:web}. Although limited in size, such a platform, called Lagrange~\cite{Viviani_LagrangeOperatingItalys_2026}, plays an important role in education, algorithm development, and the testing of control and error-mitigation techniques, contributing to the growth of local expertise and the training of the next generation of quantum scientists~\cite{Cipollini_Threewaysshare_2026}.

Despite the obstacles and challenges outlined above, Lagrange enables the definition of benchmarks grounded in realistic data-intensive applications, providing a starting point for discussing technical solutions and workarounds, while establishing a solid reference for evaluating future developments.

In this paper, a VQC-based classifier for the separation of $b$- and $c$-jets is developed using simulated open data from the LHCb experiment. The circuit is designed with a limited number of qubits in order to be executed on Lagrange. After validating the workflow on a quantum simulator, the full pipeline, including training and testing, is performed on Lagrange. This work demonstrates that the classification performance is not significantly degraded by quantum noise within the considered setup. While recent studies have demonstrated the training and evaluation of quantum models for jet tagging on quantum processing units~\cite{gluon_quark_tagging}, to the best of our knowledge, this work represents the first study using simulated data from an LHC experiment in which the complete machine-learning workflow of a VQC for jet classification, including training and testing, is performed on real quantum hardware. Methods for the evaluation of the training QPU time and for the VQC optimization are also proposed. This work can be regarded as a testbed for future applications on larger quantum processors, and a discussion on the scalability of the approach is provided.

\section{Methods}\label{sec2}

\subsection{Heavy-flavor jet reconstruction and identification at the LHCb experiment}
\label{sec:jets}

LHCb is a forward spectrometer covering the pseudorapidity range $2 < \eta < 5$ in proton--proton collisions at the LHC at CERN \cite{lhcb}. It consists of a tracking system and a particle identification system. The tracking system is composed of a vertex detector, tracking stations, and a superconducting dipole magnet. The particle identification system comprises two ring-imaging Cherenkov detectors, an electromagnetic calorimeter, a hadronic calorimeter, and a muon detection system. During data taking in 2010--2013 (Run~1) and 2015--2018 (Run~2), LHCb employed a two-level online selection (trigger) system, with a hardware-based first level followed by a software-based high-level trigger. Starting from 2022 (Run~3), the trigger is fully implemented in software. In Runs~1 and~2, LHCb collected a total integrated luminosity of about $9\,\mathrm{fb}^{-1}$ at center-of-mass energies of 7, 8, and 13~TeV.

The reconstruction of hadronic jets at LHCb proceeds as follows \cite{lhcb_zjet}. Charged particles are first reconstructed as tracks from hits in the tracking system. Energy deposits from photons and electrons in the electromagnetic calorimeter, and from hadrons in the hadronic calorimeter, are also reconstructed as clusters. These objects are combined using a particle flow algorithm \cite{CMS:2017ParticleFlow}, which selects tracks and calorimeter clusters not associated with tracks as inputs to jet clustering. Jets are then reconstructed using the anti-$k_t$ algorithm \cite{antikt} with a distance parameter\footnote{The parameter $R$ defines the characteristic radius of the jet in the pseudorapidity--azimuthal angle ($\eta$--$\phi$) plane, with $\Delta R = \sqrt{(\Delta\eta)^2 + (\Delta\phi)^2}$.} $R = 0.5$. Jet energy corrections are applied to calibrate the jet energy scale, and jet identification criteria are imposed to suppress combinatorial background \cite{Bursche:2012}.

A key ingredient in the identification of heavy-flavor jets is the secondary vertex (SV) reconstruction \cite{lhcb_tagging}. Tracks reconstructed in the vertex detector are used to identify primary vertices, corresponding to the proton--proton interaction points, and SVs displaced from them, associated with the decay of $b$- and $c$-hadrons. A jet is tagged as heavy flavor if a SV is reconstructed within the jet cone. The efficiency of this requirement is approximately 60\% for $b$ jets and 25\% for $c$ jets, with a misidentification rate of order $10^{-3}$ \cite{lhcb_tagging}, making the contamination from gluon and light-flavor ($u, d, s$) jets negligible for most applications. To further discriminate between $b$- and $c$-jets, machine-learning BDTs are employed. Ten variables related to the SV are used as inputs to two BDT classifiers: one to separate $b$- from $c$-jets, and another to further suppress light-jet contamination. The choice of the BDT working point depends on the specific analysis and is optimized to maximize physics sensitivity. 

Quantum algorithms for jet classification have already been explored within the LHCb experiment \cite{qml_tagging}, although previous studies were limited to simulations of quantum hardware. In the present study, the feasibility of deploying such algorithms on a real quantum device, namely the 5-qubit quantum computer, Lagrange, is investigated. In addition, the potential scalability of these approaches to larger quantum processors is discussed. Such systems could eventually operate beyond the regime of classical simulability, opening up the possibility of investigating whether quantum advantage can be achieved in high-energy physics classification tasks.

During the development of this work, alternative approaches based on deep neural networks have been developed \cite{lhcb_hbb}. These algorithms use on the order of 400 features related to jet substructure, significantly enhancing the discrimination between different jet flavors.

\subsection{The 5-qubit superconducting quantum computer Lagrange}
\label{sec:lagrange}

The jobs performed for the present study were dispatched to the Lagrange quantum computer, a gate-based, 5-qubit IQM Spark device~\cite{iqm_spark_2024}, hosted at Politecnico di Torino, managed by
LINKS Foundation and purchased with the contribution of INRiM. User access and quota validation for projects are implemented to allow a workflow similar to other cloud quantum computing providers~\cite{Viviani_LagrangeOperatingItalys_2026}. Lagrange can be accessed via the IQM API \cite{IQMClient}, enabling users to submit quantum circuits using Qiskit \cite{qiskit}.

As QML algorithms inherently rely on the repeated execution of parameterized quantum circuits and classical optimization loops, the underlying hardware must provide stable, parameterized gate primitives and reliable readout capabilities. The QPU of the IQM Spark is based on grounded transmon qubits. The device features five data qubits arranged in a star topology, wherein a single central qubit is physically connected to four peripheral qubits. The connections are mediated by flux-tunable couplers, allowing for dynamic control of the effective coupling strength between qubit pairs and thus better calibration. The native gate set of the IQM Spark~\cite{iqm_spark_2024} comprises single-qubit parameterized rotations (PRX), enabled by individually addressing each qubit via a dedicated microwave drive line, and two-qubit Controlled-Z (CZ) gates, realized by applying an external magnetic flux through the SQUID loop of tunable couplers \cite{Campbell:2023}. A significant advantage of the tunable coupler architecture is the active suppression of noise and crosstalk. When the QPU is idling, the couplers are tuned to a point where the parasitic ZZ-interaction between neighboring qubits is effectively zero \cite{Sung:2021}. 
This active compensation preserves high-fidelity identity operations during circuit execution. While the exact performance fluctuates depending on the calibration conditions, two-qubit CZ-gate fidelities on this architecture routinely reach $98$--$99\%$. Achieving fidelities beyond the three-nines regime ($> 99.9\%$) remains an active area of research~\cite{Acharya_Quantumerrorcorrection_2025,Marxer_999FidelitySingleQubit_2026,fidelity2,fidelity3}, as such performance is generally regarded as a prerequisite for scalable fault-tolerant quantum computation based on surface-code architectures~\cite{Fowler_Surfacecodespractical_2012}.
Calibrations on Lagrange are performed periodically. To improve stability, all jobs were submitted within a single interval between two consecutive calibrations.

State measurement on the IQM Spark is performed via dispersive readout. Each qubit is coupled to a dedicated readout resonator, which in turn is protected by an individual filter to suppress the qubit's Purcell decay rate \cite{Purcell:1946} to the environment. The signals from all readout resonators are coupled to a common probe line and amplified by a Traveling Wave Parametric Amplifier \cite{Macklin:2015} operating in a non-degenerate three-wave mixing regime. This setup allows for fast, frequency-multiplexed readout of the entire 5-qubit register within approximately $100$ ns \cite{iqm_spark_2024}. 

Since IQM Spark is a NISQ device, it is useful to describe its noise profile and its impact on QML. The relaxation time, $T_1$, dictates the rate at which qubits decay from the excited state to the ground state due to energy loss to the environment. In a QML context a short $T_1$ results in stronger amplitude damping, driving the qubit states toward the ground state and consequently suppressing part of the information encoded in the quantum state.
Typical relaxation timescales for the transmons are on the order of $T_1\sim 40-45~\mu\text{s}$, providing a sufficiently wide window to execute the shallow-to-moderate depth parameterized circuits required for our QML classifiers before amplitude damping becomes significant. 
The phase decoherence time, $T_2$, represents the time over which a qubit preserves the relative phase between the levels characterizing its superposition. QML relies heavily on the constructive and destructive interference between different computational paths to construct highly non-linear, complex decision boundaries. Dephasing noise scrambles these relative phases, effectively collapsing the coherent superposition into a classical mixed state. Consequently, in order to preserve the desired quantum state, the total time of circuit operations should be below $T_2$, which for Lagrange is of the order of about $9~\mu\text{s}$.

Several noise-mitigation strategies are discussed in Refs.~\cite{iqm_spark_2024,Cai_Quantumerrormitigation_2023}. In particular, readout error mitigation can be performed using reference measurements~\cite{Nation_ScalableMitigationMeasurement_2021}, zero-noise extrapolation can be implemented through dedicated software frameworks such as Mitiq~\cite{LaRose_Mitiqsoftwarepackage_2022}, while correlated errors can be suppressed through randomized compiling techniques, including Pauli twirling~\cite{Hashim_RandomizedCompilingScalable_2021}. These mitigation protocols were not considered in the present study, as they would introduce a significant computational overhead and increase the number of circuit executions required, but they will be considered for future studies.

\subsection{Datasets and features}

The LHCb simulation sample used in this study is available as an open dataset from the LHCb repository \cite{opendata}. It consists of simulation samples of dijets with $b\bar{b}$ and $c\bar{c}$ quarks, generated with the conditions of the Run 2 data-taking, corresponding to proton-proton collisions with $\sqrt{s}=13$ TeV collected in the years 2015-2018. The collision process, including fragmentation and hadronization of quarks, is generated with Pythia 8 \cite{pythia8}. The interaction of the particles with the LHCb detector is simulated with Geant4 \cite{geant4}. The simulation, digitization and reconstruction of the signals is performed with the LHCb software stack \cite{lhcb_reco}. Jets are reconstructed as described in Sec. \ref{sec:jets}. Jets from $b$- and $c$-quarks are selected by requiring a transverse momentum with respect to the beam axis ($p_{\mathrm{T}}$) of at least 17 GeV$/c$, and the presence of at least one SV in the jet cone defined by $\Delta R <0.5$. This way, the contamination of light jets (\emph{i.e.} generated from $u$-, $d$-, $s$-quarks or gluons) to the selected jet samples can be assumed negligible. Although the default LHCb BDT algorithm for jet tagging employs ten input features related to the SV~\cite{lhcb_tagging}, five input features are considered in this study. In fact, in the encoding scheme used (Sec. \ref{sec:methods}), each classical feature is represented by the rotation angle of a single qubit, therefore the number of qubits available in Lagrange limits the number of features.

\begin{figure}[h]
    \centering
    \includegraphics[width=0.8\textwidth]{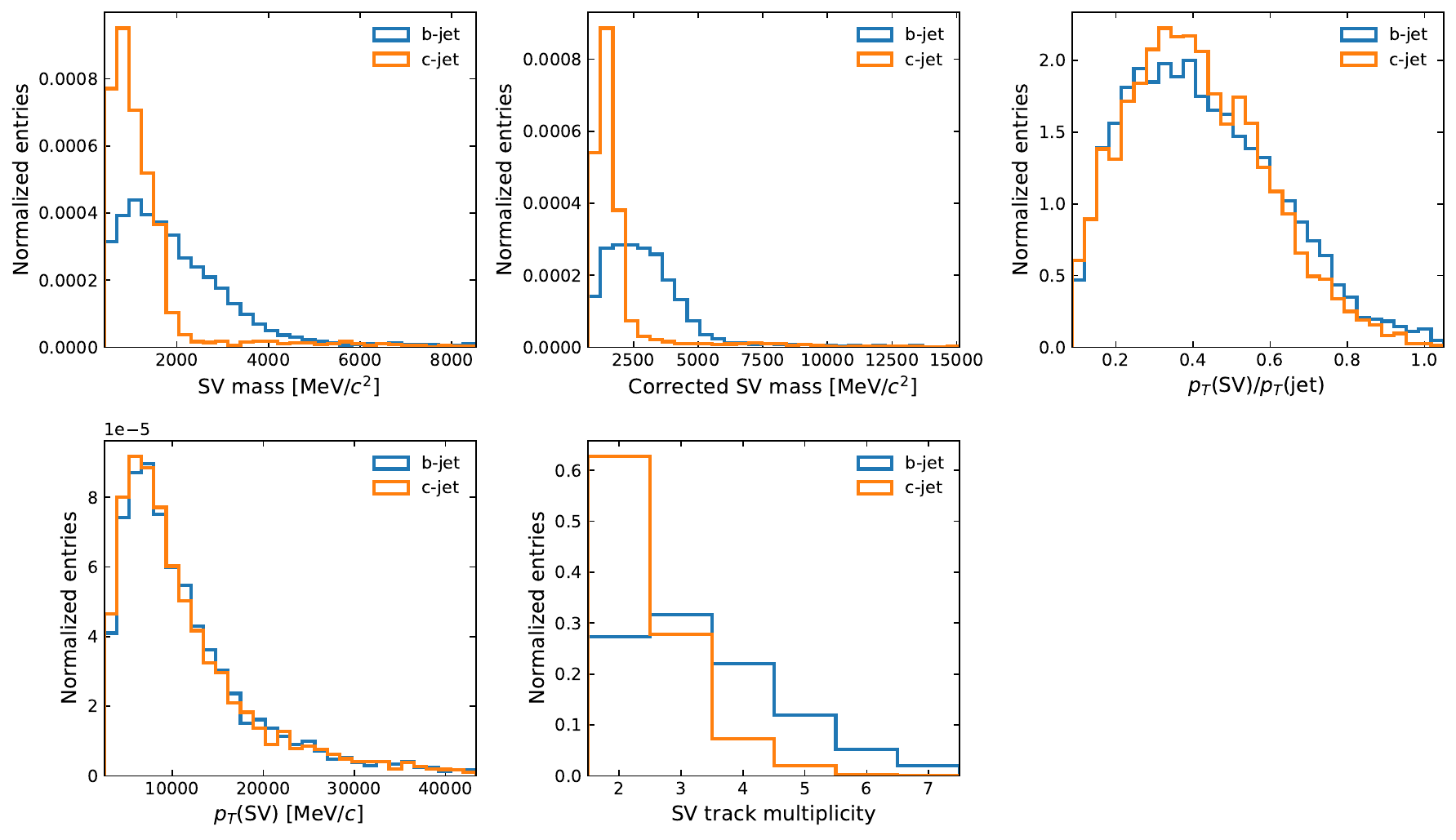}
    \caption{Distributions of the five input features used in the VQC, compared between $b$-jets and $c$-jets.}
    \label{fig:input_features}
\end{figure}

The five selected features, all associated with the secondary vertex (SV) with the highest $p_{\mathrm{T}}$ within the jet when multiple SVs are reconstructed, are:
\begin{enumerate}
\item the invariant mass of the tracks originating from the SV defined as:
\begin{equation}
m_\text{SV} = \sqrt{E^2-p^2},
\end{equation}
where $E$ and $p$ denote the total energy and momentum of the tracks forming the SV, respectively;
\item the corrected invariant mass of the SV:
\begin{equation}
m^{\text{corr}}_\text{SV} = \sqrt{m_\text{SV}^2+p^2 \cdot \mathrm{sin}^2\Theta_\text{SV}} +p \cdot \mathrm{sin}\Theta_\text{SV},
\end{equation}
where $\Theta_\text{SV}$ is the angle between the SV momentum and the SV flight direction. This mass represents the SV mass corrected for unreconstructed particles, like neutrinos;
\item the ratio between the $p_{\mathrm{T}}$ of the SV and the $p_{\mathrm{T}}$ of the jet;
\item the $p_{\mathrm{T}}$ of the SV;
\item the number of tracks that form the SV.
\end{enumerate}
With the exception of the SV $p_{\mathrm{T}}$, all the selected features are also employed by the default LHCb BDT algorithm.
The distributions of these features, compared for $b$-jets and $c$-jets, are presented in Fig. \ref{fig:input_features}.
It has been verified that a classical neural network (Sec.~\ref{sec:neural_network}) achieves comparable performance when trained using either these five features or the full set of input features employed by the default BDT algorithm.
Furthermore, smaller subsets of these five features are used when fewer qubits are considered: for $N_q$=4, where $N_q$ denotes the number of qubits, features 1 to 4 are employed, whereas for $N_q$=3, only features 1 to 3 are used. The feature numbering does not imply any ranking in terms of importance; the reduced subsets are introduced solely as an exercise to study the classifier performance with different numbers of input features and qubits.

For the classification task, the dataset of $b\bar{b}$ jets is used as the signal sample, while the dataset of $c\bar{c}$ jets is used as the background sample.
The dataset is structured such that each individual $b$- or $c$-jet constitutes a single entry. Therefore, for simulated proton--proton collision events in which dijet production occurs, approximately two entries are produced on average, corresponding to the two jets in the event. Overall, approximately 500k $b$-jet and 500k $c$-jet entries are available. However, only a subset of these entries is used for training and testing, as described in Sec. \ref{sec:hyper}.


\subsection{Quantum machine learning methods} 
\label{sec:methods} 

The classification task is performed using a VQC, a hybrid quantum--classical machine-learning algorithm based on parameterized quantum circuits. In this approach, classical input features are first encoded into the state of a quantum register through a feature map. The encoded quantum state is then processed by a trainable variational circuit, whose parameters are optimized through a classical optimization loop. The output of the circuit is finally measured and interpreted as a classification score. This work employs an approach similar to the one presented in Ref.~\cite{qml_tagging}, where $b$ and $\bar{b}$ jets were distinguished using a different set of observables.

The number of available qubits in Lagrange imposes a direct constraint on the dimensionality of the input feature space. 
The encoding scheme used is the \emph{angle embedding}, where each classical feature is represented by the rotation angle of a single qubit, allowing the $N_q$ input features to be directly embedded into an $N_q$-qubit quantum register.
For this reason, the number of input features is chosen to be equal to the number of qubits, \(N_q\). 

Before being encoded into the quantum circuit, each input feature is standardized to zero mean and unit standard deviation using \texttt{StandardScaler} \cite{scikit-learn}. The scaling parameters are determined exclusively from the training sample and subsequently applied to the test samples. The standardized features are then transformed according to $x \rightarrow \arctan(x) + \pi$, resulting in values within the interval $(\pi/2, 3\pi/2)$.

The resulting values are then embedded into the quantum state through angle encoding. For an input vector $\mathbf{x}=(x_0,\ldots,x_{N_q-1})$, where each component $x_i$ represents the rotation angle associated with a normalized feature, the feature map is defined as 
\begin{equation} |\psi_{\mathrm{enc}}\rangle = \bigotimes_{i=0}^{N_q-1} R_y(x_i) |0\rangle^{\otimes N_q}, \end{equation} 
where $R_y(x_i)$ denotes a rotation by an angle $x_i$ around the $y$-axis of the Bloch sphere. 

After the encoding stage, the quantum state is processed by a variational ansatz composed of \(N_l\) repeated layers of trainable single-qubit rotations and entangling gates. 
Three-rotation gates are implemented in order to obtain a generic unitary transformation in $\mathrm{SU}(2)$ through the Euler decomposition. The entangling layer consists of a sequence of CNOT gates arranged in a ring topology. For the five-qubit circuit considered here, the CNOT gates are applied sequentially, with control--target pairs $(q_0,q_1)$, $(q_1,q_2)$, $(q_2,q_3)$, $(q_3,q_4)$, and $(q_4,q_0)$, in this order. Thus, each qubit acts as the control of a CNOT targeting the subsequent qubit, with the last qubit $q_4$ connected back to $q_0$ to close the ring.

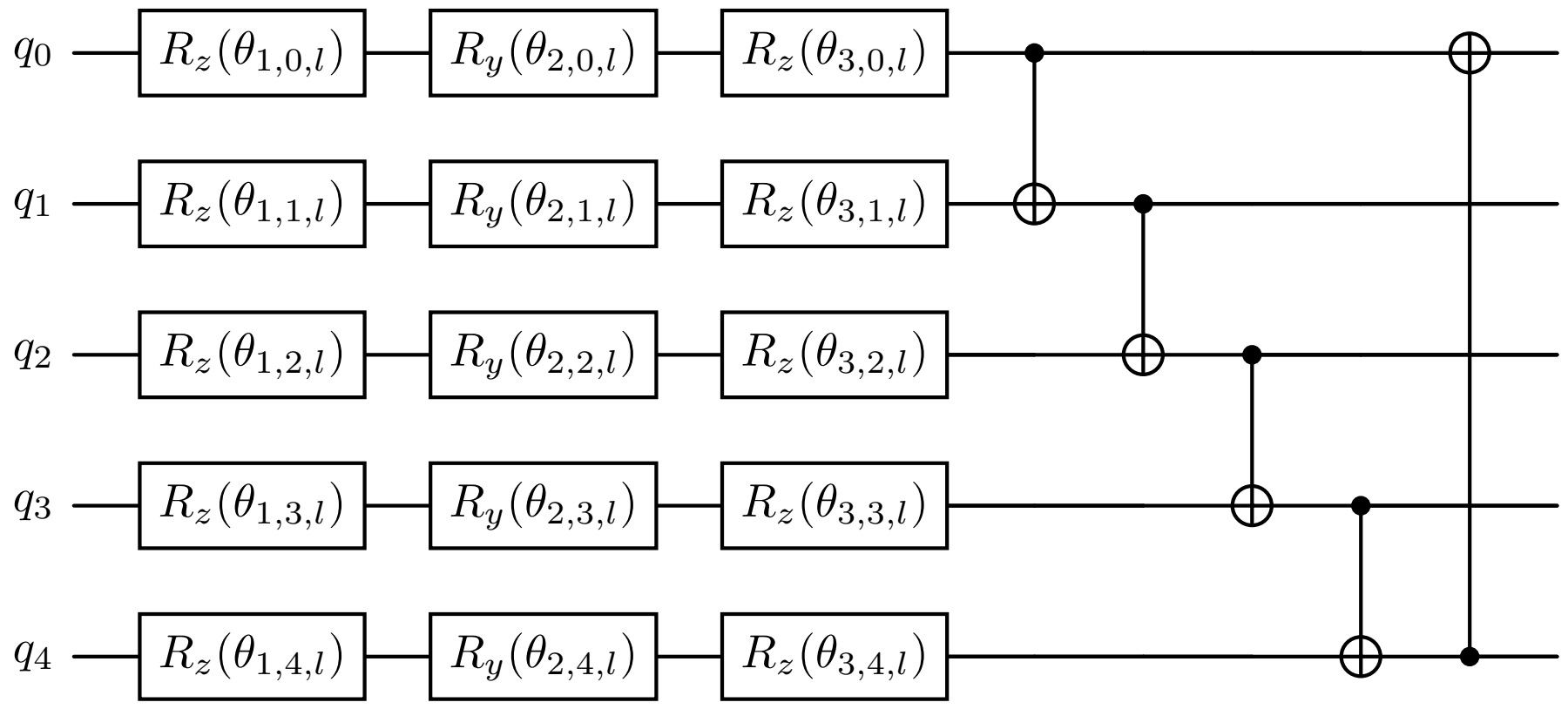
\begin{figure}[htbp] \centering \begin{quantikz} \lstick{$q_0$} & \gate{R_z(\theta_{1,0,l})} & \gate{R_y(\theta_{2,0,l})} & \gate{R_z(\theta_{3,0,l})} & \ctrl{1} & \qw & \qw & \qw & \targ{} & \qw \\ \lstick{$q_1$} & \gate{R_z(\theta_{1,1,l})} & \gate{R_y(\theta_{2,1,l})} & \gate{R_z(\theta_{3,1,l})} & \targ{} & \ctrl{1} & \qw & \qw & \qw & \qw \\ \lstick{$q_2$} & \gate{R_z(\theta_{1,2,l})} & \gate{R_y(\theta_{2,2,l})} & \gate{R_z(\theta_{3,2,l})} & \qw & \targ{} & \ctrl{1} & \qw & \qw & \qw \\ \lstick{$q_3$} & \gate{R_z(\theta_{1,3,l})} & \gate{R_y(\theta_{2,3,l})} & \gate{R_z(\theta_{3,3,l})} & \qw & \qw & \targ{} & \ctrl{1} & \qw & \qw \\ \lstick{$q_4$} & \gate{R_z(\theta_{1,4,l})} & \gate{R_y(\theta_{2,4,l})} & \gate{R_z(\theta_{3,4,l})} & \qw & \qw & \qw & \targ{} & \ctrl{-4} & \qw \end{quantikz} \caption{Structure of a single variational layer (ansatz).} \label{fig:ansatz} \end{figure}

The complete circuit can be written as \begin{equation} 
U(\mathbf{x},\boldsymbol{\theta}) = U_{\mathrm{ansatz}}(\boldsymbol{\theta}) \,U_{\mathrm{enc}}(\mathbf{x}), 
\end{equation} 
where \(\boldsymbol{\theta}\) denotes the set of trainable parameters. The variational ansatz is defined as 
\begin{equation} U_{\mathrm{ansatz}}(\boldsymbol{\theta}) = \prod_{l=1}^{N_l} \left[ U_{\mathrm{ent}} \bigotimes_{i=0}^{N_q-1} R_z(\theta_{3,i,l}) R_y(\theta_{2,i,l}) R_z(\theta_{1,i,l}) \right], 
\end{equation} 
where \(U_{\mathrm{ent}}\) represents the layer of entangling gates, $R_y$ and $R_z$ are rotations along the $y$ and $z$ axes, and \(\theta_{i,l}^{(k)}\) are trainable parameters.
The ansatz is represented in Fig. \ref{fig:ansatz}.
The classifier output is obtained by measuring the expectation value of the Pauli-\(Z\) operator acting on the first qubit ($Z_0$), 
%
which produces the circuit score 
\begin{equation} s(\mathbf{x},\boldsymbol{\theta}) = \langle \psi(\mathbf{x},\boldsymbol{\theta}) | Z_0 | \psi(\mathbf{x},\boldsymbol{\theta}) \rangle, 
\end{equation} 
where 
\begin{equation} |\psi(\mathbf{x},\boldsymbol{\theta})\rangle = U(\mathbf{x},\boldsymbol{\theta}) |0\rangle^{\otimes N_q}. 
\end{equation} 
It should be noted that the $Z_0$ measurement operator commutes with $R_z$ rotations. Consequently, the final $R_z$ rotation applied to each qubit in the final ansatz layer does not affect the measurement outcome. Moreover, the subsequent CNOT layer does not introduce any sensitivity to these rotations in the measured observable. The corresponding rotation angles can therefore be regarded as redundant variational parameters.

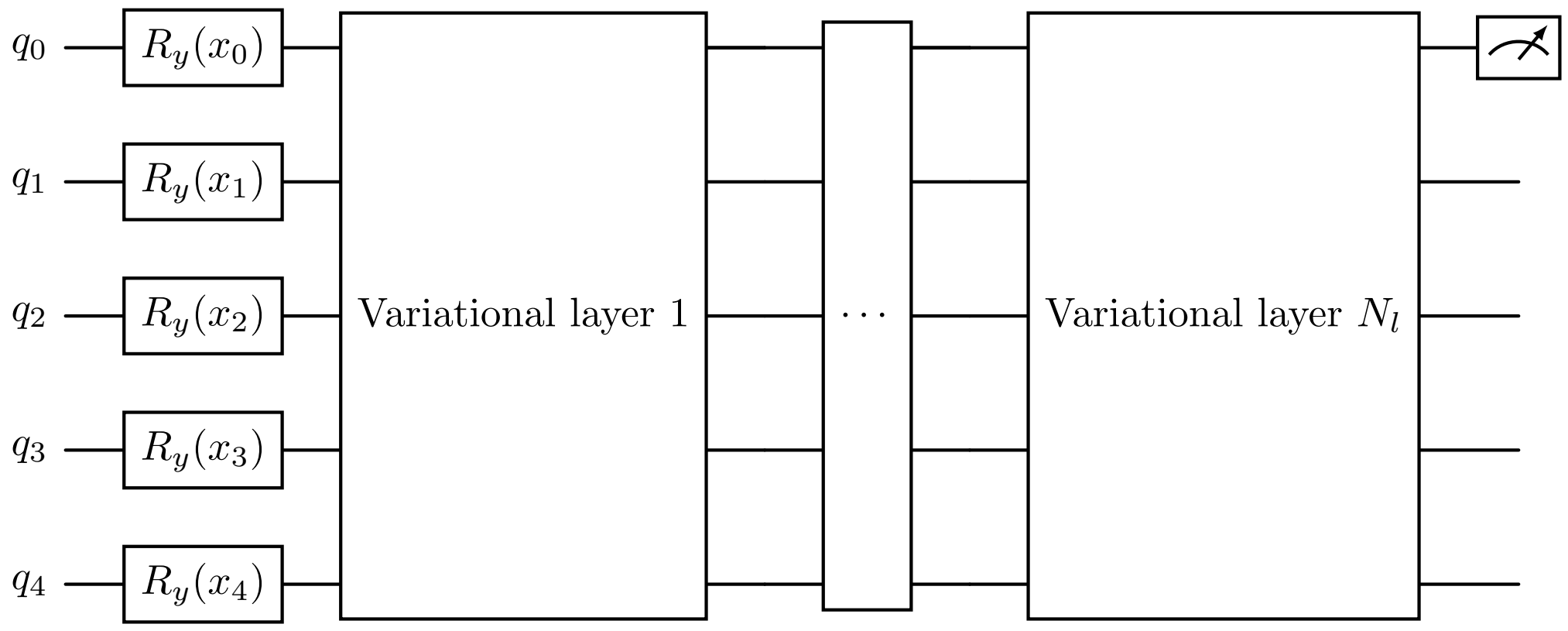
\begin{figure}[htbp] \centering \begin{quantikz} \lstick{$q_0$} & \gate{R_y(x_0)} & \gate[wires=5]{\text{Variational layer }1} & \qw & \gate[wires=5]{\cdots} & \qw & \gate[wires=5]{\text{Variational layer }N_l} & \meter{} \\ \lstick{$q_1$} & \gate{R_y(x_1)} & & & & & &  \\ \lstick{$q_2$} & \gate{R_y(x_2)} & & & & & &  \\ \lstick{$q_3$} & \gate{R_y(x_3)} & & & & & &  \\ \lstick{$q_4$} & \gate{R_y(x_4)} & & & & & &  \end{quantikz} \caption{Schematic representation of the variational quantum classifier. The input features are encoded through angle embedding using $R_y$ rotations. The encoded state is subsequently processed by $N_l$ variational layers, each composed of trainable single-qubit rotations and entangling gates. The classifier output is obtained by measuring the expectation value of the Pauli-\(Z\) operator acting on the first qubit.} \label{fig:vqc}
\end{figure}

The expectation value \(s\) is used as the output score of the classifier, and can be obtained by measuring the trained circuit several times (number of \emph{shots}, $N_{\mathrm{shots}}$) and calculating the average of the measurements on the first qubit. A schematic representation of the full quantum circuit, from feature embedding to final measurement, is shown in Fig.~\ref{fig:vqc}.

The circuit parameters are optimized through a hybrid quantum--classical training procedure as described in Ref.~\cite{gradient}. 
The optimization of the trainable parameters is performed using a classical Adam optimizer algorithm \cite{adam} with a fixed learning rate of $\lambda$. At each epoch, the output score is evaluated on the training set and the sum squared error is computed as the loss function. The gradients with respect to the circuit parameters are obtained using the parameter-shift rule:
\begin{equation}
\label{eq:parameter_shift}
\frac{\partial s(\theta_i)}{\partial \theta_i} = \frac{s\left(\theta_i + \frac{\pi}{2}\right) - s\left(\theta_i - \frac{\pi}{2}\right)}{2},
\end{equation}
Subsequently, the computed gradients are used to update the variational parameters using the Adam optimizer.


For each training event, the quantum circuit is executed on either a simulator or a quantum processing unit (QPU), producing the classifier score. The loss function is evaluated on a classical processor and the variational parameters are iteratively updated. Training is performed for a sufficient number of epochs to ensure convergence as will be discussed later. The architecture adopted in this work is specifically designed to be executable on the 5-qubit Lagrange quantum processor, while retaining sufficient expressive power for the heavy-flavor jet classification task. The dependence of the classifier performance on the circuit hyperparameters, such as the number of qubits and variational layers, is discussed in Sec.~\ref{sec:hyper}. The quantum circuits and the corresponding training workflow are implemented using Qiskit version~1.4.2 \cite{qiskit}.

\subsection{Hyperparameter optimization and timing analysis for processing on Lagrange}
\label{sec:hyper}

\subsubsection{Hyperparameters definition}
The primary objective of this study is to execute all quantum stages of the VQC training and evaluation workflow on the Lagrange quantum processor, thereby assessing the impact of quantum noise on the overall classification performance.

Given the limited amount of QPU time available on Lagrange, defined as the time during which the quantum processing unit is actively used to execute the quantum circuits, a preliminary optimization study is performed on a simulator to reduce both the execution time of individual quantum circuits and the total number of circuit evaluations while preserving adequate classification performance. The VQC architecture and training procedure are characterized by several hyperparameters that can be tuned to balance performance and computational cost.
The following hyperparameters are considered:
\begin{itemize}
\item $N_q$: the number of input features, corresponding to the number of qubits used through angle embedding;
\item $N_l$: the number of variational layers in the quantum circuit;
\item $N_{\text{shots}}$: the number of shots used for each circuit evaluation;
\item $N_{\text{train}}$: the size of the training sample, consisting of equal numbers of $b$-jets and $c$-jets randomly selected from the initial dataset;
\item $N_{\text{eval}}$: the number of entries used for evaluation;
\item $\lambda$: the learning rate;
\item $N_{\text{epochs}}$: the number of training epochs.
\end{itemize}
The total number of logical gates in the VQC, including the feature-encoding layer, is given by
\begin{equation}
N_{\mathrm{gates}} = N_q + 4N_qN_l,
\end{equation}
where the first term corresponds to the $N_q$ rotations used for data encoding, while each variational layer consists of $N_q$ CNOT gates and $3N_q$ single-qubit rotational gates. 
The number of native gates executed in Lagrange ($N_{\mathrm{gates}}^{\mathrm{Lag.}}$) can be obtained after the transpilation. 
The number of trainable parameters is
\begin{equation}
N_w = 3N_qN_l,
\end{equation}
since each rotational gate is associated with one trainable parameter, whereas the embedding rotations are determined directly by the input features and are therefore not trainable.
Given the redundancy of the last rotation applied to each qubit, as explained in Sec. \ref{sec:methods}, the effective number of weights contributing to the classification is:
\begin{equation}
N^*_w = N_w -N_q.
\end{equation}
As a starting point for the hyperparameter optimization, a \textit{reference} configuration is defined with $N_q=5$, $N_l=2$, and $N_{\mathrm{shots}}=1024$. This choice closely follows the best-performing configuration identified in our previous work, where a VQC was applied to LHCb jet data~\cite{qml_tagging}. The figure of merit used to compare the different configurations is the area under the Receiver Operating Characteristic (ROC) curve (AUC). The AUC is evaluated on the independent test sample, and, for the simulator only, its statistical uncertainty is estimated using a bootstrap procedure applied to the same sample \cite{bootstrap}. The hyperparameters defining the \textit{reference} configuration, together with the corresponding AUC, are summarized in Tab.~\ref{tab:hyperparameters}. Throughout the remainder of the paper, results obtained using the simulator are referred to as \textit{noiseless} simulations. Here, \textit{noiseless} indicates that quantum hardware noise is not included in the simulation, while statistical fluctuations arising from the finite number of shots are still present.

\subsubsection{Dependence on the training sample size and comparison with a classical neural network}
\label{sec:neural_network}

In this section the performance of the VQC as a function of the number of training events is studied using the noiseless simulator.
In order to discuss possible advantages of the quantum algorithm with respect to the classical method, the \textit{reference} configuration of the VQC is also compared to a classical neural network. 
The neural network is implemented using the \texttt{MLPClassifier} class from the scikit-learn library~\cite{scikit-learn}. The network consists of two fully connected hidden layers with four neurons each and uses the ReLU activation function. The model is trained using the Adam optimizer with an initial learning rate of $10^{-3}$ for a maximum of 10000 iterations. The input variables are standardized as for the VQC. To account for the stochastic nature of the training procedure, the neural network was trained several time using different random seeds, while keeping the training and test datasets and hyperparameters fixed. The AUC was evaluated independently for each training run. The reported AUC corresponds to the mean value across the different random seeds, while the associated uncertainty is given by their standard deviation. It has been verified that the statistical uncertainty associated to the test sample is negligible compared to this one. All other \texttt{MLPClassifier} hyperparameters are kept at their default values.
Given the five input features and a single output node, this architecture contains 49 trainable parameters, including weights and biases. The architecture is intentionally kept small so that the number of trainable parameters is of the same order of magnitude as the 25 independent variational parameters of the VQC, allowing for a more meaningful comparison between the classical and quantum models in terms of model complexity.

\begin{figure}[h]
    \centering
    \includegraphics[width=0.6\textwidth]{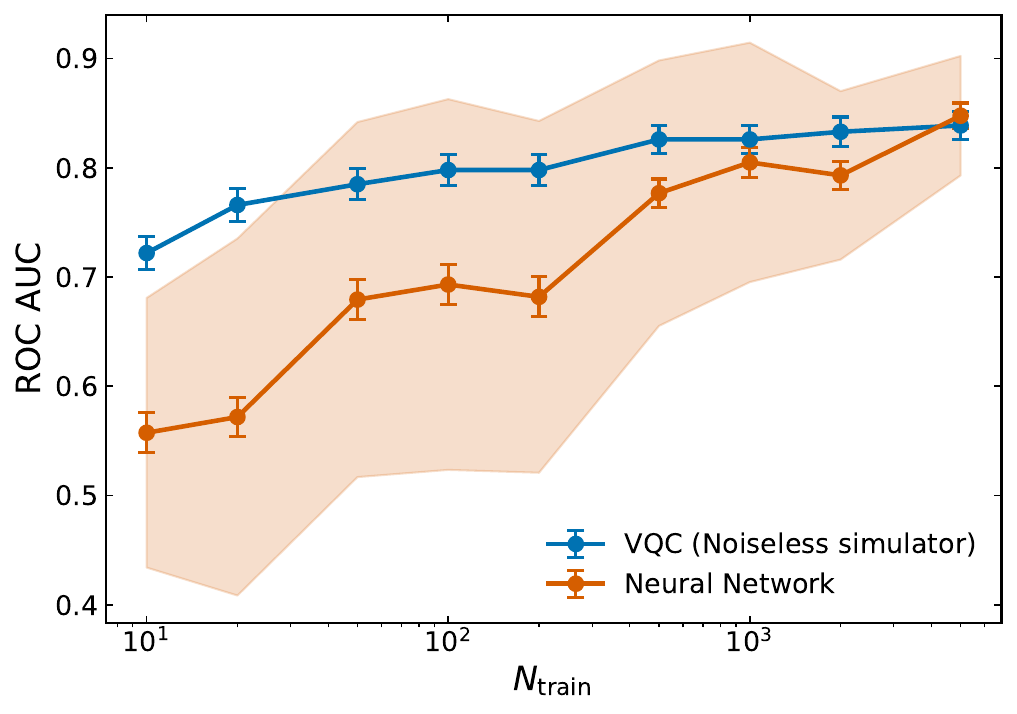}
    \caption{Area under the ROC curve as a function of the number of training entries, for the VQC with $N_q=5$ and $N_l=2$, and for a neural network with 49 tunable parameters. The shaded band associated to the neural network results represents the standard deviation of the AUC values across the runs with different random seeds, while the error bars represent the statistical uncertainty of the test sample.}
    \label{fig:nn_train}
\end{figure}

A comparison between the two models is performed by evaluating their performance as a function of the number of training samples, $N_{\mathrm{train}}$. For this study, the VQC is evaluated using a noiseless simulator. The results are shown in Fig.~\ref{fig:nn_train}. The VQC rapidly reaches a performance plateau, achieving AUC values that are consistently higher than the mean values obtained with the classical neural network. Nevertheless, the results of the two models are compatible when accounting for the variability of the neural network performance across different random seeds.

These findings suggest that the VQC can achieve competitive performance even with a relatively small training dataset, potentially requiring fewer training entries than the classical neural network to achieve comparable performance. However, the observed differences remain within the variability associated with the random seed dependence of the classical model, preventing a definitive conclusion regarding the superiority of the quantum approach.

\subsubsection{Timing analysis}
\label{sec:timing}

The QPU execution time ($t_\mathrm{QPU}$) was measured by executing the circuit on the IQM Lagrange quantum processor for different values of the relevant hyperparameters. The execution time can be extracted directly from the log files available through the Lagrange dashboard. These logs provide a detailed breakdown of the timing associated with the various stages of a job, allowing the effective time spent performing quantum operations on the hardware to be isolated and identified as the QPU execution time. Since the reported execution time is integrated over the total number of acquired shots, all measurements presented in this study are normalized to the number of shots, $N_{\mathrm{shots}}$. In addition, it was verified that the specific values of the variational parameters defining the rotation gates do not have a significant impact on the execution time.

The measurement was performed as a function of the number of qubits, $N_q$, and the total number of layers, $N_{\mathrm{l}}$. It was also verified that $t_{\mathrm{QPU}}$ is approximately proportional to the number of shots, with deviations below 10\%. These variations may be attributed to fluctuations in the operating conditions of the QPU. For this reason all timing measurements are performed with $N_{\text{shots}}=1024$, and the reported quantity is the average QPU time per shot $<\frac{t_{\mathrm{QPU}}}{N_{\text{shots}}}>$. 
This metric is studied as a function of the circuit depth $N_{\mathrm{depth}}$, defined as the number of sequential gate layers required to execute the quantum circuit, where gates acting on different qubits are assumed to be performed in parallel. For the considered ansatz, each variational layer contributes three layers of single-qubit rotations and $N_q$ layers of entangling CNOT gates, leading to
\begin{equation}
N_{\mathrm{depth}} = N_l (3+N_q)+1.
\end{equation}

\begin{figure}[h]
    \centering
    \includegraphics[width=0.6\textwidth]{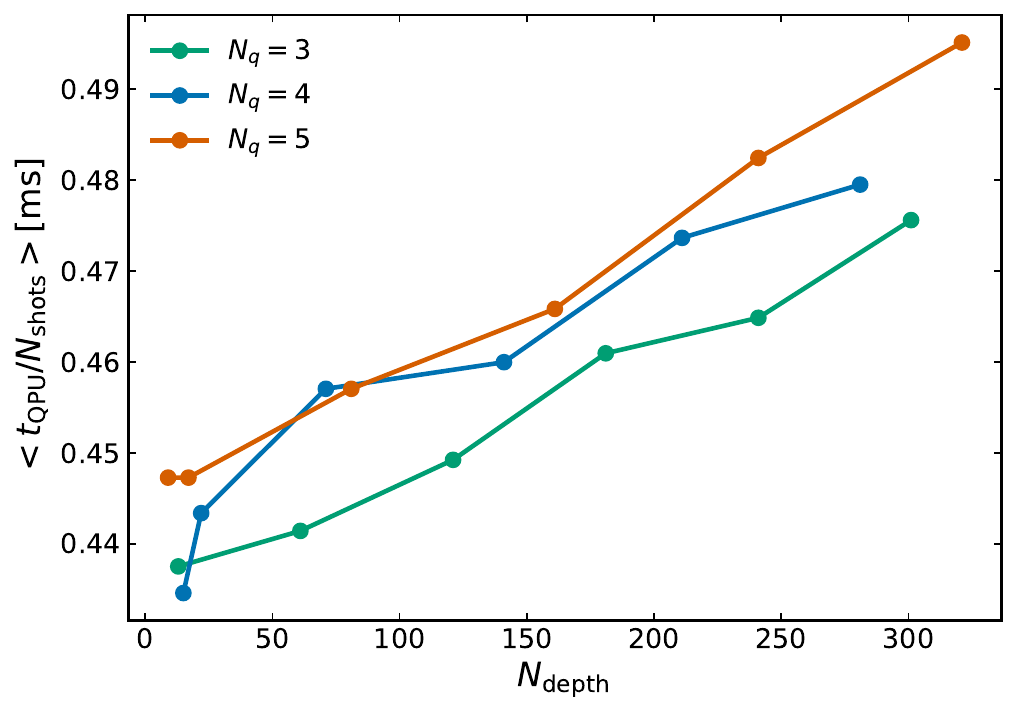}
    \caption{QPU time in Lagrange normalized by the number of shots as a function of the circuit depth and for different numbers of qubits.}
    \label{fig:t_vs_ndepth}
\end{figure}

The results of the time measurements are shown in Fig.~\ref{fig:t_vs_ndepth}.
It can be seen that the QPU execution time scales approximately linearly with the circuit depth. In contrast, only a weak dependence on the number of qubits is observed, although slightly larger execution times are measured when increasing the circuit size from $N_q=3$ to $N_q=4$.
The dependence of the execution time on the circuit complexity can be approximated by
\begin{equation}
\label{eq:time_single}
<\frac{t_{\mathrm{QPU}}}{N_{\mathrm{shots}}}>\,[\mathrm{ms}]
\simeq
0.45 + 1.56\times10^{-4}\,N_{\mathrm{depth}}.
\end{equation}
It should be noted that for each circuit execution there is a time overhead of 450 $\mu s$, which corresponds to the hardware operations that are performed to initialize and measure the qubits. The time contribution due to the execution of the gates is 0.156 $\mu$s, which is much lower than the $T_1$ and $T_2$ defined in Sec. \ref{sec:lagrange}.
It was also verified that the same proportionality law holds when considering depth given by the physical gates obtained after the transpiling instead of using the logical gates. This behaviour is expected, as the number of physical gates generally increases with the number of logical gates.

The training stage represents the dominant contribution to the total QPU usage. At each training epoch, the loss function must be evaluated by executing the circuit for all $N_{\text{train}}$ training entries. Furthermore, the gradient is estimated with the parameter-shift method of Eq. \ref{eq:parameter_shift}. 
For each trainable parameter, both upward and downward variations are applied, and the corresponding circuits are evaluated over the entire training sample. Consequently, the total number of circuit evaluations required during training is
\begin{equation}
\label{eq:n_train_circuit}
N_{\text{circuit}}^{\text{train}} =
N_{\text{epochs}}  N_{\text{train}} (1 + 2N_w).
\end{equation}
The total QPU time that is necessary for the training can be predicted with:
\begin{equation}
\label{eq:time}
t^{\mathrm{train}}_{\mathrm{QPU}} = N_{\text{circuit}}^{\text{train}} < \frac{t_{\mathrm{QPU}}}{N_{\mathrm{shots}}} > N_{\mathrm{shots}},
\end{equation}
where $ < \frac{t_{\mathrm{QPU}}}{N_{\mathrm{shots}}} >$ is obtained from Eq. \ref{eq:time_single}. 

\subsubsection{Hyperparameter optimization}

In this section, the procedure used to identify a \textit{cost-optimized} configuration of the VQC for execution on Lagrange is presented. Applying Eq.~\ref{eq:time} to the \textit{reference} configuration yields an expected training time of 4700 minutes in terms of QPU time. Since this exceeds the available QPU time on Lagrange, the hyperparameters are progressively reduced with the aim of minimizing the total training time while allowing for at most a 10\% relative degradation in the AUC with respect to the \textit{reference} configuration. 

For each set of hyperparameters, the VQC is retrained and its AUC is evaluated using noiseless simulations. The dependence of the classification performance on the different hyperparameters is then studied individually. 

Figure~\ref{fig:nn_train} shows the AUC as a function of the number of training events, $N_{\text{train}}$. 
Although the AUC already approaches a plateau for smaller training samples, $N_{\text{train}}=100$ is selected as a conservative working point. This choice reduces the potential sensitivity of the training to statistical fluctuations associated with the particular events selected in very small training samples, while retaining a substantial reduction in the required QPU time with respect to the \textit{reference} configuration.

\begin{figure}[h] 
\centering 
\includegraphics[width=0.45\textwidth]{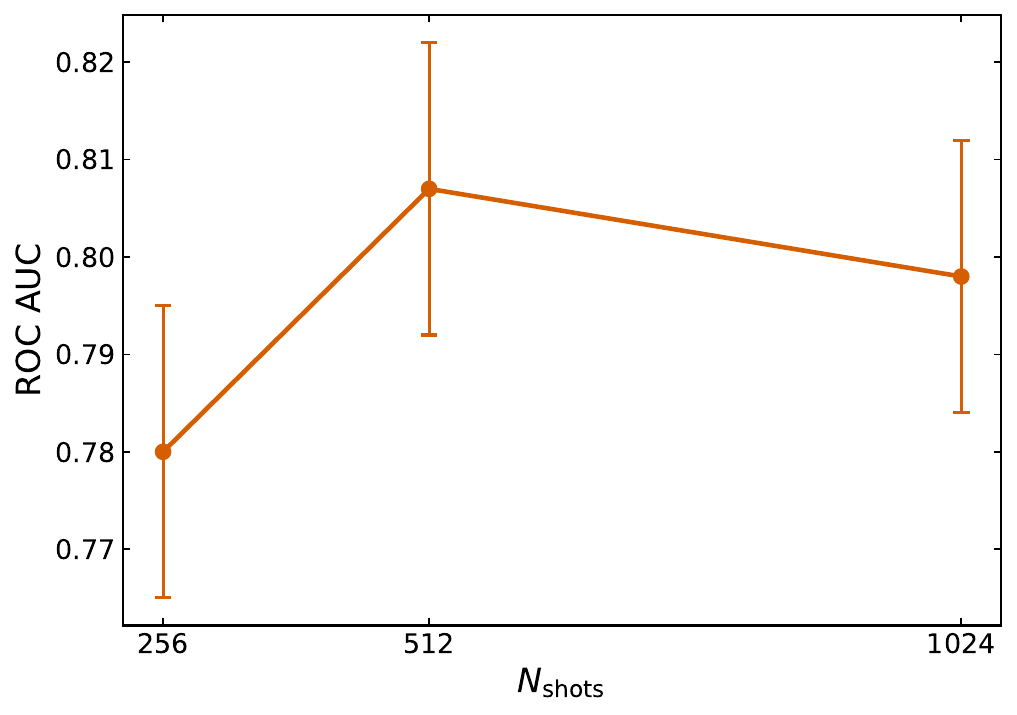} 
\includegraphics[width=0.45\textwidth]{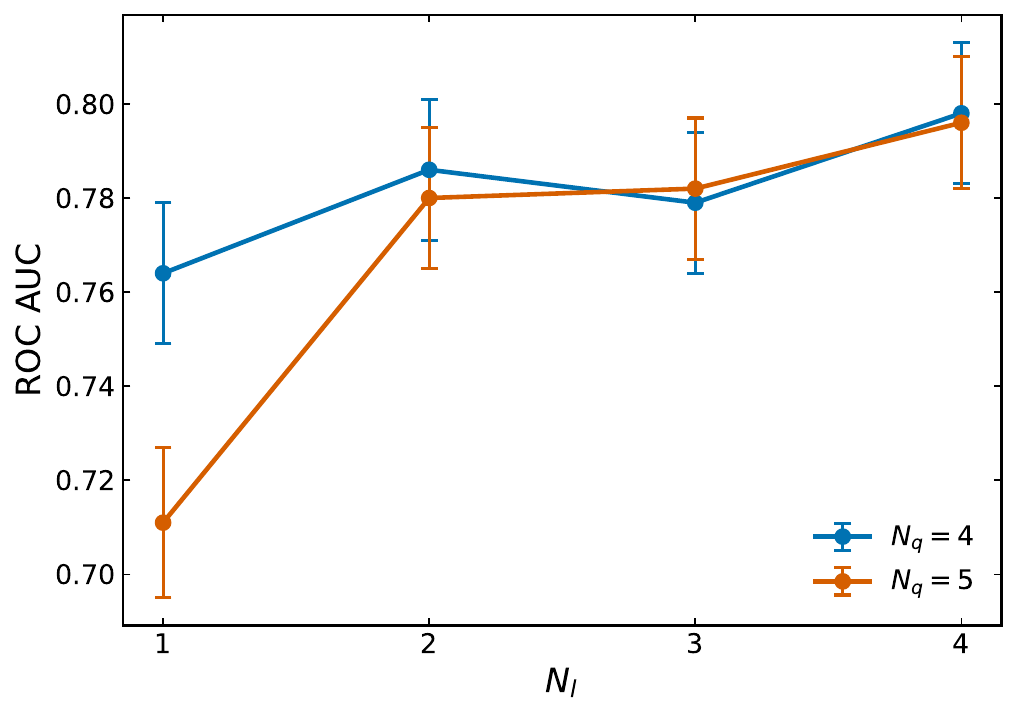}
\caption{Left: area under the ROC curve as a function of the number of shots, $N_{\mathrm{shots}}$, for $N_q=5$, $N_l=2$, and $N_{\text{train}}=100$. Right: area under the ROC curve as a function of the number of variational layers, $N_l$, for different values of $N_q$, $N_{\text{train}}=100$ and $N_{\text{shots}}=256$.} 
\label{fig:auc_vs_nshots} 
\end{figure}

As a next step, the dependence of the AUC on the number of shots, $N_{\text{shots}}$, is studied while keeping $N_q=5$ and $N_l=2$. The results are shown in Fig.~\ref{fig:auc_vs_nshots}. The AUC values obtained for $N_{\text{shots}}=256$, 512, and 1024 are compatible within their statistical uncertainties, and the lowest tested value, $N_{\text{shots}}=256$, is selected for the \textit{cost-optimized} configuration. 

Finally, the dependence of the AUC on the number of qubits, $N_q$, and variational layers, $N_l$, is investigated. Configurations with $N_q=4,5$ and $N_l=1,2,3,4$ are considered, and the corresponding results are shown in Fig.~\ref{fig:auc_vs_nshots}. The performance reaches a plateau around $N_l=2$, with no significant improvement observed when additional variational layers are introduced. For the \textit{cost-optimized} configuration, $N_q$ and $N_l$ are therefore reduced to 4 and 1, respectively. This configuration satisfies the optimization criterion, as the corresponding relative reduction in AUC remains below 10\% with respect to the \textit{reference} configuration. Specifically, the AUC decreases from 0.826 to 0.763, corresponding to a relative reduction of approximately 7.5\%. In terms of the Gini coefficient, defined as $G = 2\,\mathrm{AUC}-1$, this corresponds to a relative performance reduction of approximately 19\%~\cite{gini}.

The learning rate and the number of training epochs are fixed to $\lambda=0.1$ and $N_{\text{epochs}}=10$, respectively, after verifying the stable convergence of the optimization procedure in noiseless simulation. 

The resulting \textit{cost-optimized} configuration is summarized in Table~\ref{tab:hyperparameters}, together with the \textit{reference} configuration. Overall, the optimization leads to an AUC reduction of approximately 7.5\%, while decreasing the expected QPU training time from 4700 minutes to 48 minutes, corresponding to a reduction by a factor of approximately 100.

\begin{table}[htbp]
\centering
\caption{Reference and cost-optimized hyperparameter configurations used for the VQC training and evaluation.}
\label{tab:hyperparameters}
\begin{tabular}{l|c|c}
\hline
\textbf{Parameter} & \textbf{Reference} & \textbf{Cost-optimized} \\
\hline
$N_q$ & 5 & 4 \\
$N_l$ & 2 & 1 \\
$N_{\text{shots}}$ & 1024 & 256 \\
$N_{\text{train}}$ & 1000 & 100 \\
$N_{\text{eval}}$ & 1000 & 1000 \\
$\lambda$ & 0.1 & 0.1 \\
$N_{\text{epochs}}$ & 10 & 10 \\
\hline
$N_{\text{gates}}$ & 45 & 20 \\
$N_{\text{gates}}^{\text{Lag.}}$ & 133 & 54 \\
$N_w$ & 30 & 12 \\
$N^*_w$ & 25 & 8 \\
$N_{\text{circuit}}^{\text{train}}$ & 610000 & 25000 \\
$N_{\text{depth}}$ & 17 & 8 \\
\hline
ROC AUC (simulator) & $0.826 \pm 0.013$ & $0.763 \pm 0.015$ \\
\hline
$t^{\mathrm{train}}_{\text{QPU}}$ (expected from Eq. \ref{eq:time}) & 4700 min & 48 min \\
\hline
\end{tabular}
\end{table}

\subsection{Impact of quantum noise}
\label{sec:noise}

The studies presented in Sec.~\ref{sec:hyper} rely on noiseless simulations for the evaluation of the VQC performance. Before proceeding with the full execution on Lagrange, a dedicated test is performed to assess the impact of quantum noise.

\begin{figure}[h]
    \centering
    \includegraphics[width=0.6\textwidth]{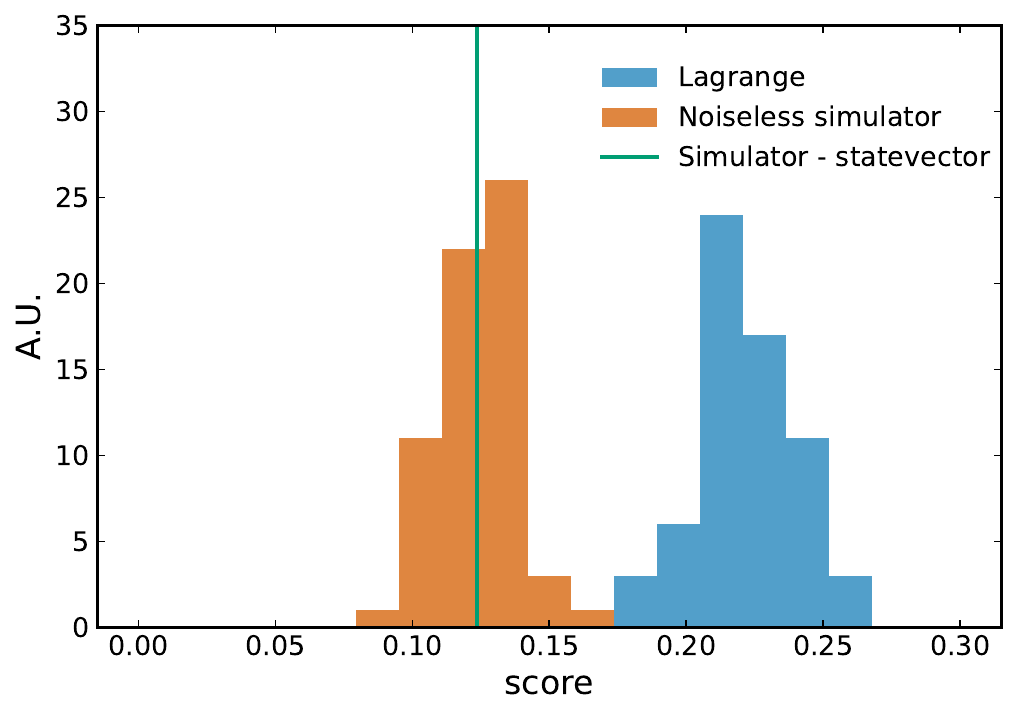}
    \caption{Score distribution for a specific event obtained by executing the same circuit with: a statevector simulation, a noiseless simulator and Lagrange. $N_q=4$, $N_l=1$ and $N_{\text{shots}}=512$ is chosen as a representative test point, and 1000 repetitions of the circuit are performed. Arbitrary units (A.U.) are used in the $y$-axis.}
    \label{fig:probability}
\end{figure}

To this end, a circuit with fixed variational parameters is evaluated repeatedly for a single event to determine the distribution of the classifier output. The evaluation is performed under three different conditions: using a noiseless statevector simulator, which provides the exact expectation value and is therefore equivalent to the infinite-shot limit; using a noiseless simulator with a finite number of shots; and using Lagrange with the same finite number of shots. For this study, $N_q=4$, $N_l=1$ and $N_{\text{shots}}=512$ is chosen as a representative test point, and 1000 repetitions of the circuit are performed. The resulting score distributions are shown in Fig.~\ref{fig:probability}.
When a finite number of shots is used, statistical fluctuations broaden the result obtained with the statevector simulation, producing an approximately Gaussian distribution around the nominal score value. Comparing this distribution with the one obtained from Lagrange reveals two additional effects attributable to quantum hardware noise: an increase in the width of the distribution and a shift of its mean value, indicating the presence of a systematic bias.

Accurately modeling the noise characteristics of a specific quantum computer, such as Lagrange, is challenging, making it difficult to reproduce them faithfully in a simulator, although several studies are currently underway.
The increase in the distribution width and the observed bias may depend on several factors, including the circuit architecture, as well as the number and type of quantum gates employed. Consequently, quantum noise can modify the output score distribution of the classifier in a complex and difficult-to-predict manner. It should be emphasized that, when the training of the VQC is performed directly on quantum hardware, these effects are intrinsically incorporated into the loss function and therefore influence the optimization process. As a result, they do not necessarily degrade the final performance, since the circuit can learn a set of parameters that partially compensates for the presence of noise. Admittedly, this compensation is not expected to occur when rotation calibration biases significantly depend on time. 
These considerations further highlight the importance of performing the training procedure on quantum hardware.
However, for small-qubit systems such as the one considered in this work, the circuit can in principle be trained using a simulator and subsequently executed on quantum hardware with a negligible loss in performance. This expectation has been verified for the \textit{reference} configuration described in Tab.~\ref{tab:hyperparameters}. The circuit is first trained using the noiseless simulator, and the resulting model is subsequently evaluated on both the simulator and Lagrange. The corresponding AUC values are found to be compatible within their statistical uncertainties. Nevertheless, the complete training on the Lagrange quantum processor, presented in the next section, was performed as a proof of concept and testbed for future applications on larger-scale quantum computers. 

\section{Results}\label{sec3}

\subsection{Algorithm performance with end-to-end training on Lagrange}

In this section, we present the results of the VQC training performed using Lagrange. Given the hybrid quantum--classical approach adopted in this work, this means that all quantum circuit evaluations are executed on Lagrange, while the classical optimization is performed separately. The classical operations, such as the computation of the loss function and the gradient from the circuit outputs, are carried out on a CPU. However, it has been verified that these operations do not significantly contribute to the overall computing time. Circuit transpilation, which constitutes another classical processing step, is performed only once at the beginning of the training procedure. Subsequently, the variational parameters are updated at each optimization step without requiring further transpilation. The hyperparameter configuration used for the training on Lagrange corresponds to the optimized setup reported in Tab.~\ref{tab:hyperparameters}. To validate the training procedure, the entire workflow is also executed using the noiseless simulator, and the results are compared with those obtained on quantum hardware. To ensure a fair comparison, the same randomly initialized variational parameters are used as the starting point for both the simulated and hardware-based training.
However, according to the considerations on quantum noise discussed in Sec.~\ref{sec:noise}, the final VQCs trained on the simulator and on the quantum processor are not expected to converge to similar parameter values. Nevertheless, it is interesting to investigate whether they can achieve comparable classification performance.  The total QPU time required for the training on Lagrange is found to be 43 minutes, in agreement at the 10\% level with the 48 minutes estimate reported in Sec.~\ref{sec:hyper}.

\begin{figure}[h] 
\centering 
\includegraphics[width=0.45\textwidth]{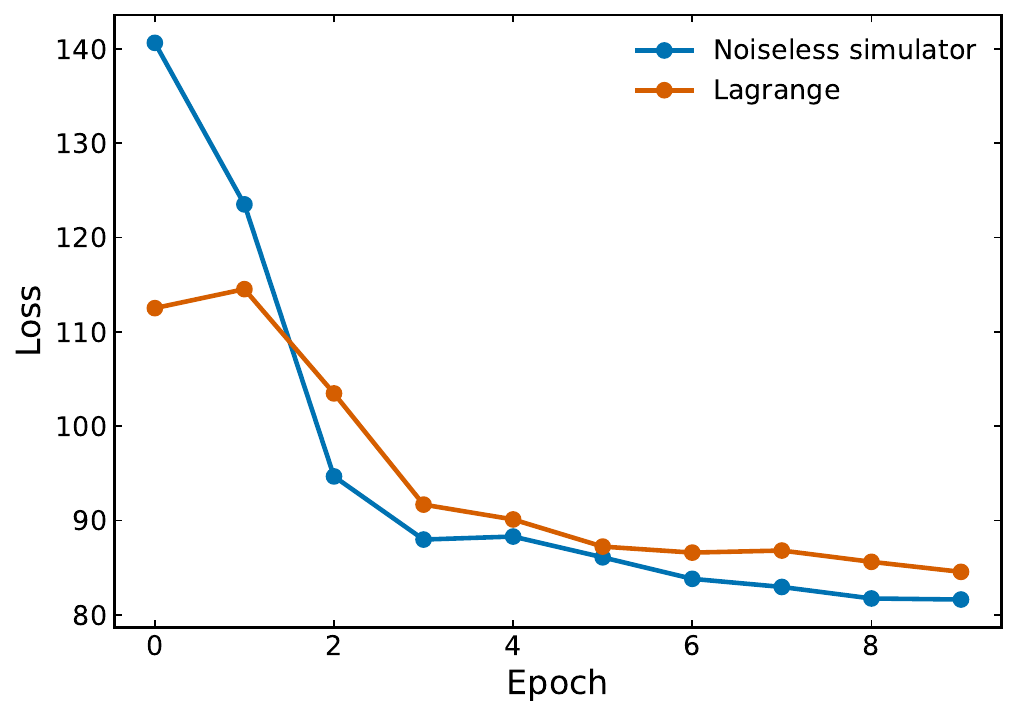} 
\includegraphics[width=0.45\textwidth]{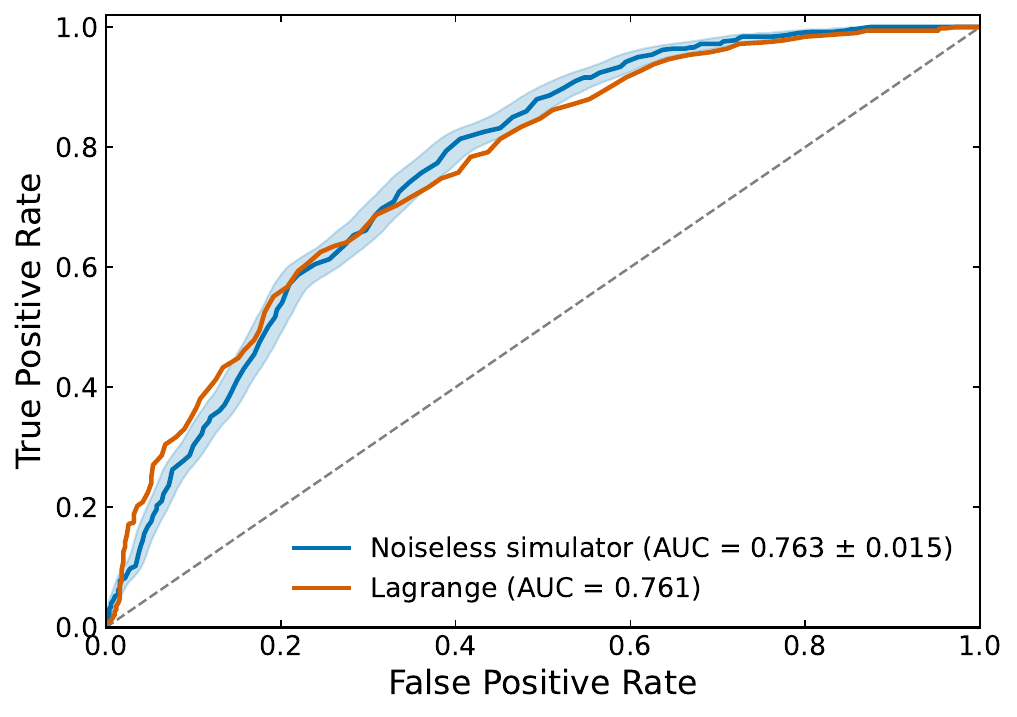}
\caption{Left: loss as a function of the training epoch for the noiseless simulator and for Lagrange. Right: ROC curves for the circuit trained and evaluated on the noiseless simulator and on Lagrange.} 
\label{fig:loss} 
\end{figure}

\begin{figure}[h] 
\centering 
\includegraphics[width=0.9\textwidth]{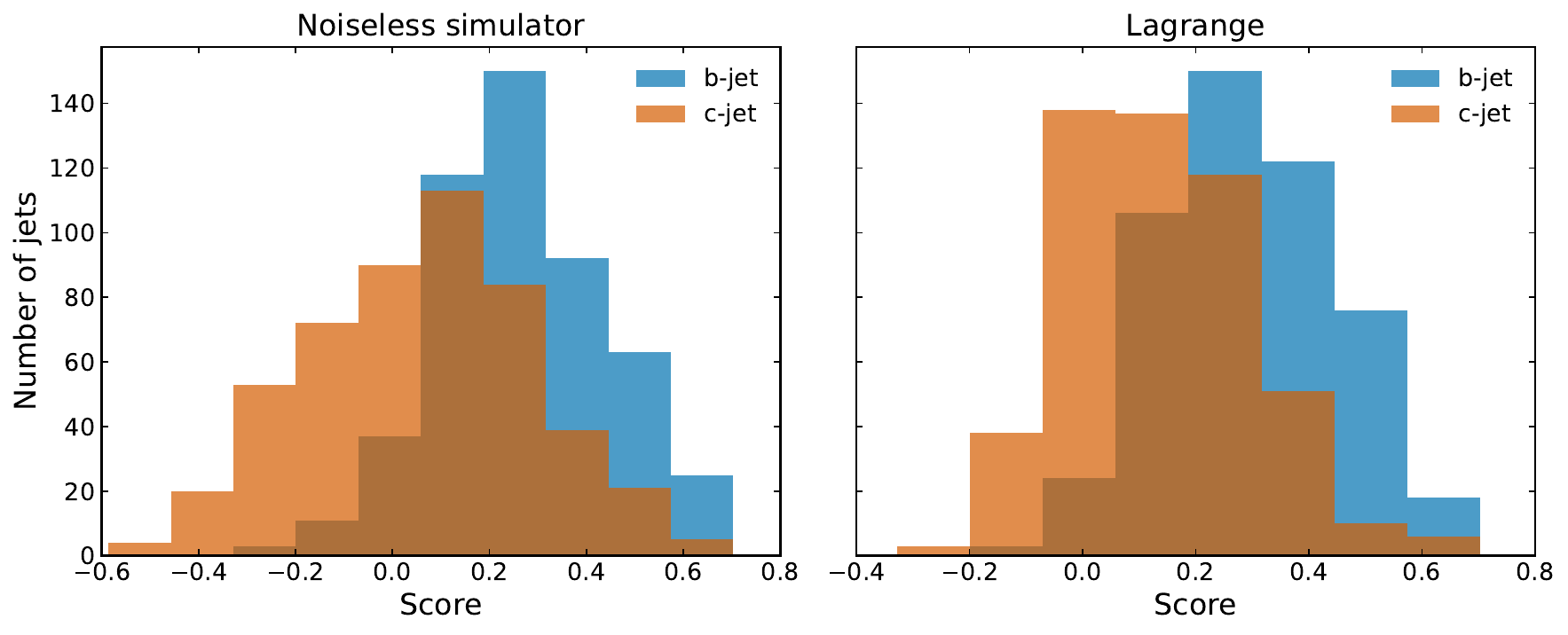} 
\caption{Output score distributions for the circuit trained and evaluated on the noiseless simulator (left) and on Lagrange (right).} 
\label{fig:score} 
\end{figure}

\begin{figure}[h] 
\centering 
\includegraphics[width=0.9\textwidth]{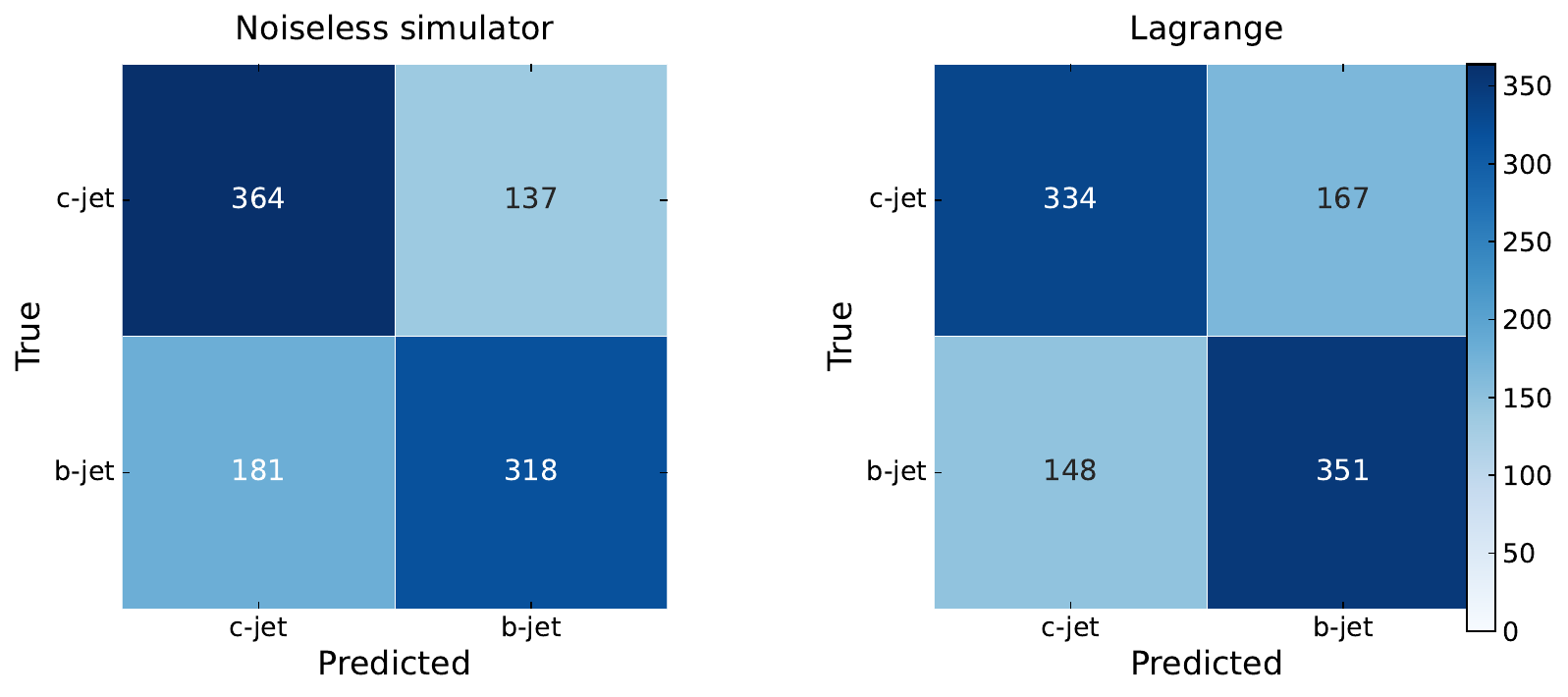} \caption{Confusion matrices for the circuit trained and evaluated on the noiseless simulator (left) and on Lagrange (right).} 
\label{fig:confusion_matrix} 
\end{figure}

The value of the loss function as a function of the training epoch is shown in Fig.~\ref{fig:loss}. The figure shows that the training performed on Lagrange reaches a convergence plateau in a manner similar to that observed for the noiseless simulator. 
The ROC curves obtained from the training on Lagrange and on the simulator are shown in Fig.~\ref{fig:loss}. Although the two curves exhibit systematic shape differences, they achieve a compatible AUC value of approximately 0.76. This result indicates that the target performance defined in Sec.~\ref{sec:hyper} is successfully achieved on Lagrange. 
The output score distributions obtained from the simulator and from Lagrange are presented in Fig.~\ref{fig:score} for both jet categories, namely $b$-jets and $c$-jets. In both cases, a partial separation between the two classes is observed. While the overall shapes of the distributions differ between the simulator and the quantum hardware, the discrimination power remains comparable. 
The confusion matrices corresponding to the two classifiers are shown in Fig.~\ref{fig:confusion_matrix}. Consistent with the ROC and score-distribution comparisons, the classification performance obtained on Lagrange is found to be similar to that achieved with the noiseless simulator. 

\section{Discussion}\label{sec4}

In this work, all quantum stages of the training and testing workflow of an optimized VQC for jet tagging have been successfully executed on the IQM Lagrange quantum processor. The observed training dynamics exhibited a clear convergence behavior, demonstrating that a complete hardware-based training loop can be performed reliably on current quantum devices. 

Furthermore, the measured execution times were found to be consistent with the proposed QPU time model presented in Sec. \ref{sec:timing} at the 10\% level, supporting its use for estimating the computational resources required for hardware-based training. It should be noted that the timing model is intended to provide an approximate estimate of the required QPU time rather than a precise prediction. In particular, it assumes an approximately linear dependence of the QPU time on the number of shots and neglects possible variations in the hardware and execution conditions between different runs. These effects can lead to deviations between the predicted and measured QPU times at the level of approximately 10\%, which is considered satisfactory for the purpose of estimating the computational cost of the different VQC configurations. The QPU time should also be distinguished from the total wall-clock time required to execute a job. The latter additionally includes contributions from job queuing, communication with the QPU, and classical preprocessing and post-processing. In the executions considered in this work, the QPU time was observed to account for approximately 20--50\% of the total wall-clock time. However, this fraction varies significantly between executions and is therefore not used to define a fixed conversion between QPU time and wall-clock time.

A particularly encouraging result is that the entire training was completed with a total QPU time of less than one hour. This demonstrates the practical feasibility of executing a complete quantum machine-learning workflow directly on quantum hardware, including the optimization of the variational parameters on the QPU itself. However, this result should be interpreted in the context of the compact \textit{cost-optimized} configuration adopted for the training on Lagrange. For jet tagging applications in high-energy physics, where larger datasets and increasingly expressive models may be required, the ability to accurately estimate and efficiently manage the required QPU resources will represent a crucial aspect for the scalability of future quantum machine-learning applications.

The comparison between the classifier trained on real quantum hardware and the corresponding classifier trained on an ideal simulator reveals visible differences in their outputs. This result is expected, since noise is inherently included in the hardware training procedure and therefore influences the optimization process. 
Despite these differences, the overall performances of the two classifiers are similar and consistent with the expectations derived from the noise-aware analysis. This observation is particularly relevant for QML applications, as it suggests that the optimization process is sufficiently robust to produce meaningful results even in the presence of realistic hardware noise. For the jet-tagging task considered in this work, the hardware implementation therefore achieves a level of performance compatible with both the simulation-based prediction and the expected impact of noise.

Beyond the specific results obtained on the 5-qubit Lagrange processor, this work introduces a general methodology for executing a complete quantum machine-learning training workflow on real hardware. The same methodology could, in principle, be applied to future quantum processors with larger numbers of qubits and improved performance, such as the 54-qubit processor being deployed at CINECA. However, its practical scalability remains to be investigated. Extending the present approach to larger datasets and circuits may require a substantially larger number of circuit executions and shots, while deeper and more highly parametrized variational circuits may also encounter optimization challenges, including barren plateaus. These aspects will require dedicated studies as larger quantum processors become available.



\begin{figure}[t]
    \centering
    \includegraphics[width=0.6\linewidth]{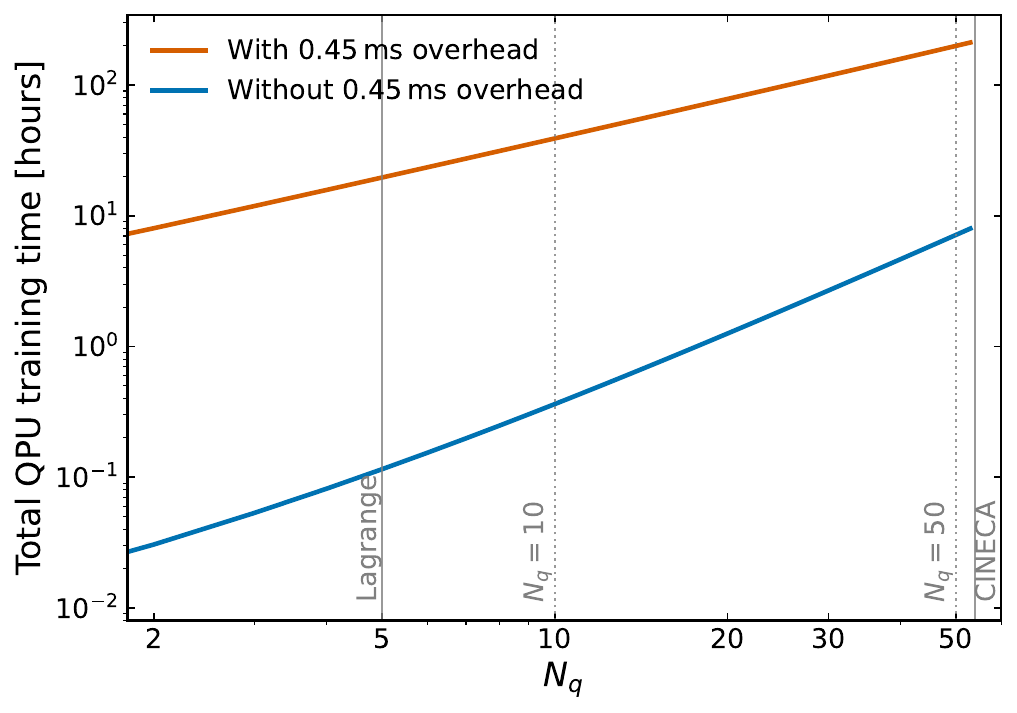}
    \caption{
    Predicted total QPU training time as a function of the number of qubits for a variational quantum classifier with $N_l=2$, $N_\text{shots}=256$, $N_{\mathrm{train}}=1000$, and $N_{\mathrm{epochs}}=10$. The prediction is obtained assuming the same dependence of QPU execution time on the number of circuit layers measured on the IQM Lagrange system. The orange curve includes the constant overhead term of Eq.~\ref{eq:time_single} ($0.45\,\mathrm{ms}$), while the blue curve shows the idealized scenario in which this contribution is neglected.
    The vertical lines at $N_q=5$ and $N_q=54$ indicate the number of qubits of the Lagrange and CINECA quantum processors, respectively.
    }
    \label{fig:qpu_scaling}
\end{figure}

As an illustrative example, a VQC architecture with $N_l=2$, $N_\text{shots}=256$, $N_{\mathrm{train}}=1000$, and $N_{\mathrm{epochs}}=10$ can be considered, while assuming the same relationship between QPU execution time and circuit complexity measured on Lagrange. Figure~\ref{fig:qpu_scaling} shows the predicted total training time as a function of the number of qubits. Under these assumptions, the required QPU time increases from approximately 40 hours for $N_q=10$ to about 200 hours for $N_q=50$.

Although these execution times are significantly larger than those observed for the present jet-tagging study, they remain potentially feasible within dedicated allocations of future quantum computing resources. Interestingly, the scaling analysis indicates that a large fraction of the total execution time originates from the constant contribution of $0.45\,\mathrm{ms}$ appearing in Eq.~\ref{eq:time_single} rather than from the gate-dependent term itself. In the regime in which this contribution exceeds the gate execution time, the overall scaling with the number of qubits is linear. This finding suggests that improvements in the hardware preparation of the quantum system could substantially reduce the overall cost of training. In the idealized scenario in which this constant overhead becomes negligible compared to the gate-dependent contribution, the required training time is dramatically reduced, as illustrated in Figure~\ref{fig:qpu_scaling}. In this regime, the training time exhibits a quadratic dependence on the number of qubits. Such improvements would significantly enhance the practicality of hardware-based quantum machine learning and could make larger-scale jet-tagging applications accessible on next-generation quantum processors.

It should also be noted that \(N_{\mathrm{train}} = 1000\) may not be sufficient to prevent overtraining when a large number of features, and consequently qubits, are employed, despite the ability of VQCs to reach their performance plateau with a relatively smaller training sample compared to classical methods. On the other hand, most of the training time is spent evaluating the gradient through \(2N_w\) circuit executions using the parameter-shift rule. Therefore, the training cost could be significantly reduced by adopting more efficient gradient-estimation techniques, such as Simultaneous Perturbation Stochastic Approximation \cite{spsa}, or by employing surrogate models based on classical neural networks to approximate the circuit output \cite{surrogate}.

\section{Conclusion}\label{sec5}

A complete hardware-based training and testing workflow for a variational quantum classifier applied to jet tagging has been demonstrated on the IQM Lagrange quantum processor. The training process showed clear convergence, while the measured execution times were successfully described by the proposed QPU time model. Differences between hardware-trained and simulator-trained classifiers were observed as a consequence of quantum noise affecting the optimization procedure. Nevertheless, the obtained performances remained similar to the simulator ones and compatible with expectations, confirming the feasibility of training quantum machine-learning models directly on current noisy quantum hardware. The developed methodology provides a practical framework for estimating the resources required by future hardware-based quantum machine-learning applications. Further improvements on the training procedure are possible, like using more efficient gradient-estimation techniques to reduce the training time. Scaling studies indicate that reducing hardware overheads will be crucial for enabling larger and more expressive quantum models on next-generation quantum processors.

\backmatter

\section*{Acknowledgements}

The authors acknowledge access to the IQM quantum computer Lagrange~\cite{Viviani_LagrangeOperatingItalys_2026}, hosted at Politecnico di Torino (Turin, Italy), managed by the LINKS Foundation and jointly acquired by Politecnico di Torino, the Fondazione LINKS, and Istituto Nazionale di Ricerca Metrologica (INRiM), which was used to perform the quantum experiments reported in this work.
The authors would like to thank the LHCb Collaboration for providing the open simulated data used in this study. In particular, we are grateful to the members of the LHCb Data Processing and Analysis Project and the LHCb Simulation Project for the production, validation, and maintenance of the simulated samples used in this work. Special thanks to Ahmed Abdelmotteleb for his thorough review and valuable suggestions, which significantly improved this work.

\section*{Funding}
We acknowledge the financial support from PNRIC ‘Q-SUD’ (1\_16868501004\_0001741), Action 1.1.2 ‘Q-SUD’ (1741\_QIIR112\_00051).

\section*{Data and code availability}
The simulated data used in this work are publicly available through the CERN Open Data Portal~\cite{opendata}. The source code used to produce the results presented in this work is available from the corresponding author upon reasonable request.

\section*{Declaration of generative AI and AI-assisted technologies}
Portions of this work were developed with the assistance of ChatGPT and Copilot in accordance with the CC-BY 4.0 license. After using this tool/service, the authors reviewed and edited the content as needed and take full responsibility for the content of the published article.









\bibliography{sn-bibliography}

\end{document}